\documentclass[final,3p, twocolumn]{elsarticle}
 \usepackage{lineno}

\usepackage{graphics}
\usepackage{slashed}
\usepackage{multicol}

 \def\tstrut{\vrule height2.5ex depth0pt width0pt} 

\journal{Nuclear Physics A}

\begin{document}

\begin{frontmatter}

\title{Heavy Quark Spin Symmetry and SU(3)-Flavour Partners of the X(3872)}

\author[IFIC]{C. Hidalgo-Duque}
\author[IFIC]{J. Nieves}
\author[PARIS]{M. Pav\'on Valderrama}

\address[IFIC]{Instituto de F\'{i}sica Corpuscular (IFIC), Centro Mixto CSIC-Universidad de Valencia, Institutos de Investigaci\'on de Paterna, Aptd. 22085, E-46071 Valencia, Spain.}
\address[PARIS]{Institut de Physique Nucl\'eaire, Universit\'e Paris-Sud,
    IN2P3/CNRS, F-91406 Orsay Cedex, France}

\begin{abstract}
In this work, an Effective Field Theory (EFT) incorporating light SU(3)-flavour
and heavy quark spin symmetry is used to describe charmed
meson-antimeson bound states.
At Lowest Order (LO), this means that only contact range interactions among the heavy meson
and antimeson fields are involved.
Besides, the isospin violating decays of the $X(3872)$ will be used
to constrain the interaction between the $D$ and a $\bar{D}^*$ mesons
in the isovector channel.
Finally, assuming that the $X(3915)$ and $Y(4140)$ resonances are $D^*\bar{D}^*$
and $D_s^*\bar{D}_s^*$ molecular states, we can determine the four Low Energy Constants (LECs) of the EFT that appear at LO and, therefore, the full spectrum
of molecular states with isospin $I=0$, $\frac{1}{2}$ and $1$.
\end{abstract}

\begin{keyword}
Heavy Quark Spin and Flavour Symmetries, Hidden charm molecules, XYZ states.
\end{keyword}

\end{frontmatter}

\section{Introduction}

Since the discovery of the X(3872) resonance \cite{Choi:2003ue}, many theoretical approaches have been used trying to describe it. Even though it contains a $c \bar{c}$ pair, it does not seem to fit in the charmonium spectrum.
Because of that disagreement, other more exotic proposals have been made. Among these, the interpretation of the X(3872) as a hadronic molecule is the most likely so far. Within this assumption, the X(3872) would be a bound state of both a charm meson and antimeson.
However, this hypothesis strongly depends on the $J^{PC}$ quantum numbers of the resonance, which have not been experimentally established. They are either $1^{++}$ or $2^{-+}$ (analysis done in Ref.~\cite{Hanhart:2011tn}),
of which only $1^{++}$ is compatible with a low-lying S-wave hadronic molecule.

Apart from the X(3872), many other experimental hidden charm resonances have been observed: the XYZ states. Many of these states can be new candidates for heavy meson-antimeson molecules.

These resonances are near their threshold. This means that the meson and the antimeson are not so close to be sensitive to find the details of the interaction at short distances. So, both of them are preserving their individuality and will not probe the specific details of the short range interaction responsible of their binding. Hence, a scale separation exists and the interaction between these two mesons can be described by different EFT's \cite{Fleming:2007rp, Valderrama:2012jv}. Besides, the presence of a heavy quark in the heavy mesons imposes that our EFT should be consistent with Heavy Quark Spin Symmetry (HQSS), which implies certain constraints to the heavy meson-antimeson interaction \cite{AlFiky:2005jd}. On the other hand, the light quark content of the heavy mesons ($q = u,d,s$ in this work) imposes SU(3) flavour symmetry and, because of that, molecular states should be classified into SU(3) multiplets.

As a consequence of the symmetries discussed (HQSS and SU(3)) flavour symmetry) only four parameters are enough to describe the molecular states at leading order in the EFT we use. That is, we need four data points to predict the full molecular spectrum. For this purpose we will assume the molecular nature of certain XYZ states such as the X(3872), X(3915) and Y(4140) and the fourth assumption will be derived from the analysis of the isospin violating branching ratio of the X(3872) decays into $J/\psi \omega$ and $J/\psi \rho$.

\section{The EFT Description at Lowest Order}
\label{sec:EFT-LO}

In this section, the EFT used in this work to describe the heavy meson molecules is briefly presented.
The EFT description must involve pions and heavy meson/antimeson fields
and local interactions among these degrees of freedom that
are compatible with the known low energy symmetries,
most notably HQSS and chiral symmetry.
Nevertheless, according to \cite{Nieves:2012tt}: pion exchanges are weaker
than naively expected and only
enter as a perturbation at subleading orders. Similarly, coupled channel  effects turn out to be also sufficiently small to be ignored at LO.
Hence, at ${\rm LO}$,
the EFT consists of heavy mesons and antimesons interacting
through a contact range potential, similar to the one in \cite{AlFiky:2005jd}.
Then, once we have determined our potential, the wave functions and observables will be calculated in the standard
quantum mechanical fashion.
For example, we can generate bound states by iterating the EFT potential
in the Schr\"odinger / Lippmann-Schwinger equation, as previously done in \cite{Nieves:2012tt}. See details in \cite{HidalgoDuque:2012pq}.

\begin{table*}[top]
\parbox{.56\linewidth}{
\begin{center}
\begin{tabular}{||c|c|c|c||}
\hline \hline \tstrut
$J^{PC}$ & $\rm H\bar{H}$ 
& $E$ $(\Lambda =0.5$ GeV) & $E$ $(\Lambda =$ 1 GeV) \\
\hline \tstrut
$0^{++}$ & $ D\bar{D}$ 
& $3709 \pm 10$ & $3715^{+12}_{-15} $ \\
\hline\tstrut
$1^{++}$ & $ D^*\bar{D}$
& Input & Input \\
$1^{+-}$ & $ D^*\bar{D}$ 
& $3815 \pm 17$ & $3821^{+23}_{-26}$ \\
\hline\tstrut
$0^{++}$ & $ D^*\bar{D}^*$& Input
& Input\\ 
$1^{+-}$ & $ D^*\bar{D}^*$ 
& $3955\pm 17$ & $3958^{+24}_{-27}$ \\
$2^{++}$ & $ D^*\bar{D}^*$ 
& $4013^{\dagger\dagger}_{-9} $ & $4013^{\dagger\dagger}_{-12} $ \\
\hline \hline
\end{tabular}
\end{center}
\caption{
Predicted masses (in MeV) of the SU(2) isoscalar  HQSS partners
of the $X(3872)$ resonance for two different values of the Gaussian
cutoff.
Errors in the predicted masses are obtained by adding in quadratures 
the uncertainties stemming  from the two sources of systematic errors
discussed at the end in Subsect.~\ref{sec:partners}. $\dagger\dagger:$ see discussion in \cite{HidalgoDuque:2012pq} }
\label{tab:hqs-partners-isoscalar}
}
\hfill
\parbox{.42\linewidth}{
\begin{center}
\begin{tabular}{||c|c|c|c||}
\hline \hline \tstrut
$J^{PC}$ & $\rm H\bar{H}$ 
& $E$ $(\Lambda =0.5$ GeV) & $E$ $(\Lambda =$ 1 GeV) \\
\hline \tstrut
$0^{+}$ & $ D_s^+\bar D^-$
& $3835.8^{+2.3}_{-3.9}$ & $3837.7^{+0.4}_{-4.3} $  \\
\hline \tstrut
$1^{+}$ & $D_s \bar{D}^*$
& $3949 \pm 13$ & $3957^{+14}_{-19}$  \\
\hline \tstrut
$0^{+}$ & $ D_s^*\bar{D}^*$ &
$4056\pm 22$
& $4061^{+29}_{-33}$ \\
$1^{+}$ & $ D_s^*\bar{D}^*$ 
& $4091^{+13}_{-14}$ & $4097^{+15}_{-20}$\\
$2^{+}$ & $ D_s^*\bar{D}^*$ 
& $-$ & $-$ \\
\hline \hline
\end{tabular}
\end{center}
\caption{
Predicted masses (in MeV) of the 
isospinor ($I=\frac{1}{2}$)   HQSS partners of the $X(3872)$ resonance,
for two different values of the Gaussian cutoff. The meaning of the
quoted errors in the table 
is the same as in Table \ref{tab:hqs-partners-isoscalar}.} 
\label{tab:hqs-partners-isospinor}
}
\end{table*}

\section{Isospin Symmetry Violation in the $X(3872)$}
\label{sec:isobreaking}
The first step is to solve the bound state equation for the $X(3872)$.
We consider that the $X(3872)$ is a $D\bar{D}^*$ molecule with quantum numbers
$J^{PC} = 1^{++}$, where we distinguish between the neutral
($D^0\bar{D}^{*0}$) and charged ($D^{+} D^{*-}$) components
of the wave function.
That is, there are two channels in the bound state equation.
And we will regularize the ${\rm LO}$ potential (here, and in the following cases when not specified) with
a Gaussian regulator function with a cut-off
$\Lambda$ varying in the $0.5-1.0\,{\rm GeV}$ range.

Moreover, the Belle collaboration reported the
decays of the $X(3872)$ into the (isoscalar) $J / \Psi \pi^+ \pi^- \pi^0$
and the (isovector) $J / \Psi \, \pi^+ \pi^-$
channels. The latest measurements yield to the isospin violating ratio \cite{Choi:2011fc}:
\begin{eqnarray}
\mathcal{B}_{X} = \frac{\Gamma (X(3872) \to J / \Psi \, \pi^+ \pi^- \pi^0)}
{\Gamma (X(3872) \to J / \Psi \, \pi^+ \pi^-)} = 0.8 \pm 0.3  \nonumber \\
\end{eqnarray}
which is difficult to accomodate from the theoretical point of view.

In this work, we will assume that this isospin violation is caused by the isospin breaking
generated by the mass difference of the neutral ($D^0 \bar{D}^{0*}$)
and charged ($D^+ D^{*-}$) channels in the
$X(3872)$, which would not
have  a definite isospin. In this picture, at short $D\bar D^*$ distances, the
$X(3872)$ would be a linear combination of $I=0$ and $I=1$
components whilst the $J / \Psi\, 3\pi$ and $J / \Psi\, 2\pi$ decays would be described by an isospin invariant coupling via an intermediate $\rho$ and $\omega$ meson (as suggested in~\cite{Gamermann:2009uq}).

The analysis performed by
Hanhart et al. in~\cite{Hanhart:2011tn}, leads to the following branching ratio
\begin{eqnarray}
R_X =
\frac{\mathcal{M}(X \to J / \Psi \,\rho)}{\mathcal{M}(X \to J / \Psi \, \omega)}
= 0.26^{+0.08}_{-0.05} \, , \label{eq:ratio}
\end{eqnarray}
where $\mathcal{B}_X$ is translated into a ratio of the decay amplitudes of the $X(3872)$  instead of its corresponding decay widths.
In our model, this ratio $R_X$ can be rewritten as (details in \cite{HidalgoDuque:2012pq})
\begin{eqnarray}
R_X = \frac{g_{\rho}}{g_{\omega}}\,
\frac{\hat{\Psi}_{X0} - \hat{\Psi}_{X1}}{\hat{\Psi}_{X0} + \hat{\Psi}_{X1}} \, .
\end{eqnarray}
being $g_{V} = \mathcal{M}_{V}[ D\bar{D}^{*} (I = 0,1) \rightarrow J/\Psi ~V]$ with $V = \rho / \omega$
and $\hat{\Psi}_{X0}$ y $\hat{\Psi}_{X1}$ an average of the neutral and charged $D^{*} \bar{D}$ wave function components in the vicinities of the origin.

So, with the experimental determination of $R_X$ and the binding energy of
the $X(3872)$ we could determine the contact range potential that binds the $X(3872)$ if we knew the $g_{\rho}/g_{\omega}$ ratio,.
But the $g_{\rho}/g_{\omega}$ ratio can be determined from the SU(3) relation $g_{\rho} - g_{\omega} = -\sqrt{2}\,g_{\phi} \, ,$
and, as the OZI (the strange  quark pair creation is suppressed) rule implies $g_{\rho}, g_{\omega} \gg g_{\phi} \, ,$
that is, we are ignoring the X(3872) decay into $J/\psi \phi$.
Thus, we can approximately consider $g_{\rho}/g_{\omega} \simeq 1$, so the
$R_X$ ratio in our model is
\begin{eqnarray}
R_X =
\frac{\hat{\Psi}_{X0} - \hat{\Psi}_{X1}}{\hat{\Psi}_{X0} + \hat{\Psi}_{X1}} \, .
\end{eqnarray}
which only depends on the wave function components in the vicinities of the origin.

\section{The SU(3) and HQSS Partners of the $X(3872)$}
\label{sec:partners}

If we determine the value of the counterterms of
the ${\rm LO}$ EFT,
we will be able to calculate the location of the molecular partners of the $X(3872)$.

There are four unknown LEC's.
We fix two of them from the location of the $X(3872)$ resonance and
its isospin breaking branching ratio, as explained in the previous section.
The remaining two require the identification of two partners of the $X(3872)$: the $X(3915)$ as a $0^{++}$
$D^*\bar{D}^*$ molecule and the $Y(4140)$ as a $0^{++}$ $D_s^*\bar{D}_s^*$ molecule, guided by its apparently
dominant decay into $J / \Psi \phi$.

Apart from the $R_{X}$ errors, there is an extra error source which has to be taken into account: the approximate nature of HQSS. In this EFT, we expand the QCD lagrangian into powers of $(\Lambda_{QCD} / m_{Q})$ so that
\begin{eqnarray}
V^{\rm LO}_{(m_Q = m_c)} = V^{\rm LO}_{(m_Q \to \infty)}\,
(1 \pm \frac{\Lambda_{QCD}}{m_c}) \, ,
\end{eqnarray}
Taking an approximate value of $m_{c}\simeq  1.5$ GeV
for the charm quark mass  and
$\Lambda_{\rm QCD} \sim 200$ MeV, we should expect a
15\% violation of HQSS for the LO contact range potentials. Since the two error sources are independent, the
total error will be computed by adding in quadratures the partial errors.

\subsection{The SU(2) Isoscalar ($I=0$) Partners}
\label{sec:isos}

In this sector the hidden strange components are ignored. 
We do not take into account particle coupled channel effects 
as they are subleading unless the mass gap between the neutral and charged channels is similar in size to the binding energy
in the isospin symmetric limit. Thus, in the $1^{++}$ and $2^{++}$ channels, we are using a coupled channel potential.
The analysis of this sector was previously done in Ref.~\cite{Nieves:2012tt} without including strangeness into the analysis (SU(2) light quark flavour symmetry) and, as there is almost no difference in the resonances predicted, we conclude that the effect of the isospin violation and the inclusion of the quark $s$ must be small.
The spectrum of molecular states is presented in
Table~\ref{tab:hqs-partners-isoscalar}.

\begin{table*} [top]
\parbox{.5\linewidth}{
\begin{center}
\begin{tabular}{||c|c|c|c||}
\hline \hline \tstrut
$J^{PC}$ & $\rm H\bar{H}$ 
& $E$ $(\Lambda =0.5$ GeV) & $E$ $(\Lambda =$ 1 GeV) \\
\hline \tstrut
$0^{++}$ & $ D^+\bar{D}^0$ 
& $3732.5^{+2.0}_{-3.9}$ & $3734.3^{+0.1}_{-3.6} $  \\
\hline \tstrut
$1^{++}$ & $ D^*\bar{D}$ 
& $-$ & $-$  \\
$1^{+-}$ & $ D^*\bar{D}$ 
& $3848^{+12}_{-11}$ & $3857^{+13}_{-18}$  \\
\hline \tstrut
$0^{++}$ & $ D^*\bar{D}^*$ & $3953 \pm 22$
& $3960^{+28}_{-32}$  \\
$1^{+-}$ & $ D^*\bar{D}^*$ 
& $3988\pm 13$ & $3995^{+14}_{-19}$ \\
$2^{++}$ & $ D^*\bar{D}^*$ 
& $-$ & $-$ \\
\hline \hline
\end{tabular}
\end{center}
\caption{Predicted masses (in MeV) of the SU(2) isovector  HQSS partners
of the $X(3872)$ resonance for two different values of the Gaussian cutoff. 
The meaning of the quoted errors  in the
table is the same as in Table \ref{tab:hqs-partners-isoscalar}.}
\label{tab:hqs-partners-isovector}
}
\hspace{0.5cm}
\parbox{.5\linewidth}{
\begin{center}
\begin{tabular}{||c|c|c|c||}
\hline \hline \tstrut
$J^{PC}$ & $\rm H\bar{H}$
& $E$ $(\Lambda =0.5$ GeV) & $E$ $(\Lambda =$ 1 GeV) \\
\hline \tstrut
$0^{++}$ & $ D_s\bar{D}_s$ 
& $3924^{+6}_{-7}$ & $3928^{+7}_{-11} $   \\
\hline \tstrut
$1^{++}$ & $ D_s^*\bar{D}_s$ 
& $-$ & $-$  \\
$1^{+-}$ & $ D_s^*\bar{D}_s$ 
& $4035\pm 15$ & $4040^{+20}_{-24}$  \\
\hline \tstrut
$0^{++}$ & $ D_s^*\bar{D}_s^*$ 
 & Input
& Input \\
$1^{+-}$ & $ D_s^*\bar{D}_s^*$ 
& $4177\pm 16 $ & $4180^{+21}_{-24}$ \\
$2^{++}$ & $ D_s^*\bar{D}_s^*$ 
& $-$ & $-$ \\
\hline \hline
\end{tabular}
\end{center}
\caption{
Predicted masses (in MeV) of the  hidden strange  isoscalar HQSS partners
of the $X(3872)$ resonance for two different values of the Gaussian cutoff. 
 The meaning of the
quoted errors is the same as in Table \ref{tab:hqs-partners-isoscalar}. }  
\label{tab:hqs-partners-strange}
}
\end{table*}

\subsection{The Isospinor ($I=\frac{1}{2}$) Partners}

In this sector, the C-parity of the molecules is not well defined, as they are not bound states of a heavy meson
and its antimeson.
However, the formalism is identical to the one in the previous case, except for the $1^{+}$ $D_s \bar D^*$ and $D
\bar D_s^*$ molecules. The $D_s \bar D^*$ and $D \bar D^*_s$ thresholds are separated by only $2\,{\rm MeV}$
and require a coupled channel treatment. In this case, we obtain just one single bound state.
This resonance and its other isospinor partners are shown
in Table \ref{tab:hqs-partners-isospinor}, where we have considered
only the strangeness one states (it would be the same for strangeness $S = -1$).

\subsection{The Isovector ($I=1$) Partners}

The potential in the isovector and isospinor sector is the same (except in the $1^{++}$ and $2^{++}$ molecules 
owing to the isospin violation). For this reason, the spectrum in the isovector sector would be similar to the isospinor one.
The four molecular states obtained are listed in Table \ref{tab:hqs-partners-isovector}.
The other two possible states correspond to the isovector partners of
the $X(3872)$ and $X(4012)$ resonances where a single bound state is obtained.

\subsection{The Hidden Strange Partners}

In this sector, the resonances must also contain a  $s \bar{s}$ quark-antiquark pair.
The potential derived for this latter case is the arithmetic
mean of the isoscalar and isovector one.
So if there was a bound state in the isoscalar and isovector
sector, there would probably exist its hidden strange partner.
On the other hand, considering that the $X(3872)$ and $X(4012)$ molecules have no isovector partners, it is very likely that there will not be hidden strange partners either with those quantum numbers.
Therefore, the four states obtained are listed in Table \ref{tab:hqs-partners-strange}.

\section{Conclusions}

As a summary, we have established the location of 15 molecular partners
of the  $X(3915)$, $Y(4140)$ and $X(3872)$ states using HQSS and SU(3) flavour symmetry, see Tables
\ref{tab:hqs-partners-isoscalar}-\ref{tab:hqs-partners-strange}.
These predictions have a series of uncertainties,
being the most important the approximate nature of HQSS. 
According to the estimations of these uncertainties, the spectrum of these states should be stable but their location can vary up to a few tens of ${\rm MeV}$
in certain cases.

However, the family of $D^{(*)}\bar D^{(*)}$ states we theorize
depends on the assumptions we made: the molecular description of the $X(3872)$, $X(3915)$
and $Y(4140)$ resonances. 
Whilst there is a consensus on the molecular nature of the X(3872) resonance, and its existence is 
well established, the situation for the $X(3915)$ and $Y(4140)$ are more dubious.
Thus not all the predictions are probable alike.
Predictions derived from the $X(3872)$ are supposed to be more solid
than those depending on the $X(3915)$, which in turn are less
hypothetical that the ones obtained from the $Y(4140)$.
In this sense, as mentioned in Ref.~\cite{Nieves:2012tt} too, 
the $2^{++}$ ${\rm D^* \bar{D}^*}$ isoscalar partner of
the $X(3872)$ is still the most reliable prediction
of the present work.
So, if any new analysis of the XYZ resonances provides us with a better molecular candidate than one of our assumptions,
it can easily be included in this scheme. Thus, we will be able to replace one of the doubtful molecular states we have assumed with this new resonance so the new predictions are more robust.
On the other way, if eventually any of the predictions we have established is detected, it could be a proof to the molecular nature of these resonances.

\begin{center}
\section*{ACKNOWLEDGMENTS}
\end{center}
This research was supported by DGI and FEDER funds, under contract
FIS2011-28853-C02-02, and the Spanish Consolider-Ingenio 2010 Programme
CPAN (CSD2007-00042),  by Generalitat Valenciana under contract
PROMETEO/20090090 and by the EU HadronPhysics2 project,
grant agreement no. 227431. 
\\

\end{document}